\documentclass[aps,prl,twocolumn,groupedaddress]{revtex4}
\usepackage{natbib}
\usepackage{graphicx}
\usepackage{dcolumn}
\usepackage{amsmath,amsthm}
\usepackage{amsfonts}
\usepackage{amssymb}
\usepackage{bbold}
\usepackage{times}
\usepackage{mathrsfs}
\usepackage{color}
\usepackage{gensymb}
\usepackage{times}
\usepackage[]{graphicx}
\usepackage{bm}
\DeclareGraphicsExtensions{.PNG,.jpg,.pdf,.gif,.eps}
\usepackage{amsmath,amssymb,amsthm,amsfonts,eucal}
\usepackage[english]{babel}

\begin{document}
\title{Optimality of contraction-driven crawling}
\author{P. Recho${}^{1,2}$, J.-F. Joanny ${}^{1,3}$ and L. Truskinovsky${}^2$ }
\affiliation{ ${}^1$ Physico-Chimie Curie UMR 168, Institut Curie, 5 Rue Pierre et Marie Curie, 75005 Paris, France\\ ${}^2$ LMS,  CNRS-UMR  7649, Ecole Polytechnique, Route de Saclay, 91128 Palaiseau,  France\\ ${}^3$ Ecole Superieure de Physique et de Chimie Industrielles de la Ville de Paris - ParisTech, Paris, France}

\date{\today}

\begin{abstract}
 We study a model of cell motility where the condition of optimal trade-off between performance and metabolic cost can be made precise.  In this model a steadily crawling fragment is represented by a layer of active gel placed on a frictional surface and driven by contraction only. We find analytically the distribution of contractile elements (pullers) ensuring that the efficiency of self-propulsion is maximal. We then show that   natural assumptions about advection and diffusion of pullers produce a distribution that is remarkably close to the optimal one and is qualitatively similar to the one observed in experiments on fish keratocytes.
\end{abstract}

\maketitle

Although the idea of optimal trade-offs in biology is rather natural, the examples
supporting optimality at the quantitative level are few. A prominent case is the
demonstration that the structure of transport networks  minimizes energy
consumption at a fixed material cost \cite{Hu2013}. Each
validation of this type is of considerable interest  as a  step towards
the general understanding of homeostasis \cite{Mortimer2011,
Chubukov2012, Szekely2013}. In this Letter we present a new example
of cost-performance trade-off associated with contraction-driven cell motility.
We show that the observed distribution of molecular motors in crawling fish
keratocytes \cite{Verkhovsky1999} is close to the optimal one and explain the mechanism behind this optimality.

The efficiency of self-propulsion in viscous environments has been a subject of
intense studies since the pioneering work of G.I. Taylor \cite{Taylor1951}.
Optimal strategies for various Stokes swimmers were identified under a tacit
assumption that the  organism is able to perform the desired shape changes
\cite{Lauga2009, Alouges2008}.
Similar reasoning has been  applied to crawling on frictional surfaces
where the optimal propulsion can be induced by actuators prescribing spatially and
temporarily correlated compression and stretch \cite{Desimone2012}.  Such
approach, however, does not reveal the physical mechanisms ensuring optimal
actuation and it remains unclear whether actual crawling cells can follow the
optimal strategy.

To address these questions we choose the simplest case of
keratocytes whose motility initiation is largely contraction driven. Experiments show that
contractile elements ('pullers') are narrowly localized at the trailing edge
\cite{Verkhovsky1999, Yam2007, Recho2013a} and our goal will be to check whether such
 configuration is optimal in terms of the trade-off between
the Stokes performance and the energetic cost of active force generation and whether it is compatible with the physical model of  motor redistribution. In our analysis we neglect the presence of 'pushers' because
active treadmilling does not play an important role at the stage of motility initiation \cite{Carlsson2011,Recho2013}. Several comprehensive
computational  studies of crawling taking treadmilling into account but not addressing the question of optimality are available in the literature \cite{Rubinstein2009}.

We model a steadily crawling cell fragment using a one-dimensional version of
the active gel theory \cite{Kruse2005, Julicher2007, Marchetti2013, Recho2013a}.
The cytoskeleton is interpreted as an infinitely compressible viscous fluid, adhesion
is represented by a frictional interaction with a rigid substrate and the cortex
is assumed to impose a fixed size on the moving cell.  Using these simplifying
assumptions  and choosing parameters in the biological range we show that the
stationary distribution of motors is close to the optimal one.

Consider a one-dimensional layer of active gel placed on a frictional rigid
background \cite{Kruse2005, Julicher2007}. The balance of forces takes the form
$$\partial_x\sigma=\xi v, $$ where $\sigma(x,t)$ is the stress, $v(x,t)$ is the
velocity and $\xi$ is the  friction coefficient.  We model the gel as a viscous
fluid subjected to active contractile stresses $\tau \geq 0$.   We can then write
 $$\sigma=\eta \partial_xv+ \tau,$$ where $\eta$ denotes  viscosity.

The regime of interest is when the cell moves with a constant velocity $V$ while
maintaining a  length $L$ fixed by the  cortex.  We therefore look for the
configuration $\sigma(y)$ and $v(y)$ depending on the  moving coordinate $y=(x-
Vt)/L$  and satisfying the boundary conditions $v(\pm 1/2)=V$ and $\sigma(\pm
1/2)=\sigma_0$, where $\sigma_0$ is the reaction stress due to the length
constraint. Without loss of generality, we assume $V\geq 0$.

The task is to find the distribution of active stresses $\tau(y)$ ensuring optimal
efficiency
$$\Lambda=P/H,$$
where $P$ is the \emph{functional} power and $H$ is the metabolic cost per unit time. In the
absence of an explicit cargo, we assume that the useful work is the translocation
of the cell as a whole against frictional resistance. Therefore we write  $$P=\xi
V^2 L$$  as in the theory of Stokes swimmers \cite{Lighthill1952}. The rate of free
energy consumption can be written as a sum   $H=H^{*}+H^{**}$  where  $$H^{*}=-\int_{-1/2}^{1/2}\tau \partial_yvdy$$ is the power exerted by the
active  stress $\tau(y)$  on the environment and  $H^{**}$
is  the  cost
of the \emph{maintenance} of the force generating machinery \cite{Hill1938}.

First we suppose that the physical mechanism of force generation is \emph{unknown}
and pose the problem of finding the function $\tau(y)\geq 0 $ maximizing $\Lambda$
at a given value of $H^{**}$.  We also prescribe average value of the contractile
stress
 \begin{equation}\label{velstress2}
 \bar{\tau}=\int_{-1/2}^{1/2}\tau(y)dy,
\end{equation}
which is equivalent to fixing the total number of motors at constant cell length.
It will be convenient to use non-dimensional variables  $\sigma/\bar{\tau}$,
$x /\sqrt{\eta/\xi}$ and $t/(\eta/ \bar{\tau})$  without changing the notations.
In dimensionless variables the
stress distribution can be written as
\begin{equation}\label{velstress1}
 \sigma(y)=\sigma_0\frac{\cosh\left( \mathcal{L}y\right)}
 {\cosh(\mathcal{L}/2)}+\mathcal{L}\int_{-1/2}^{1/2}\Psi(z,y)\tau(z) dz,
 \end{equation}
 where  $$\Psi=  \frac{\sinh(\mathcal{L}(\frac{1}{2}-y)) \sinh(\mathcal{L}
 (\frac{1}{2}+z))}{\sinh(\mathcal{L})}-\theta(z-y)\sinh(\mathcal{L}(z-y)),$$
 $\theta$ is the Heaviside function and  $\mathcal{L}=L\sqrt{\xi/\eta}$ is a parameter of the problem.
  The constants $V$ and $\sigma_0$ can be found explicitly (
 cf. \cite{Carlsson2011})
\begin{equation}\label{velstress}
\begin{array}{c}
V=-\frac{\mathcal{L}}{2}\int_{-\frac{1}{2}}^{\frac{1}
{2}}\frac{\sinh(\mathcal{L}y)}{\sinh(\frac{\mathcal{L}}{2})}\Delta\tau dy\text{,
}\\
\sigma_0-\bar{\sigma}_0=\frac{\mathcal{L}}{2}\int_{-\frac{1}{2}}^{\frac{1}
{2}}\frac{\cosh(\mathcal{L}y)}{\sinh(\frac{\mathcal{L}}{2})}\Delta\tau dy,
\end{array}
\end{equation}
where $\Delta\tau=\tau(y)-1$ is the spatially inhomogeneous component of the distribution of
active \emph{dipoles} and $\bar{\sigma}_0=1$  is the pre-stress induced by its
constant part. The first of these formulas states that contraction-induced
crawling is due entirely to spatial \emph{asymmetry}: this is an analog of the famous
Scallop
Theorem \cite{Purcell1977}. Since the total force dipole produced by the system
 is $\mathcal{L}\int_{-1/2}^{1/2}yv(y)dy=\sigma_0-\bar{\sigma}_0$  the second formula in (\ref{velstress}) states that the inhomogeneity of motor distribution is  also at the origin of a force dipole applied by the cell to the background \cite{Schwarz2002}. Using \cite{Verkhovsky1999, Kruse2005, Julicher2007, Barnhart2011}  we obtain $\bar{\tau} \sim   10^3 Pa $, $\xi \sim  2 \times 10^{16}Pa \cdot m^{-2} \cdot s$, $ \eta \sim  10^{5} \times Pa \cdot s$ and $ L\sim 20\times10^{-6} m$ which gives $\mathcal{L}\sim 10.$

\begin{figure}[!h]
\begin{center}
\includegraphics[scale=0.45]{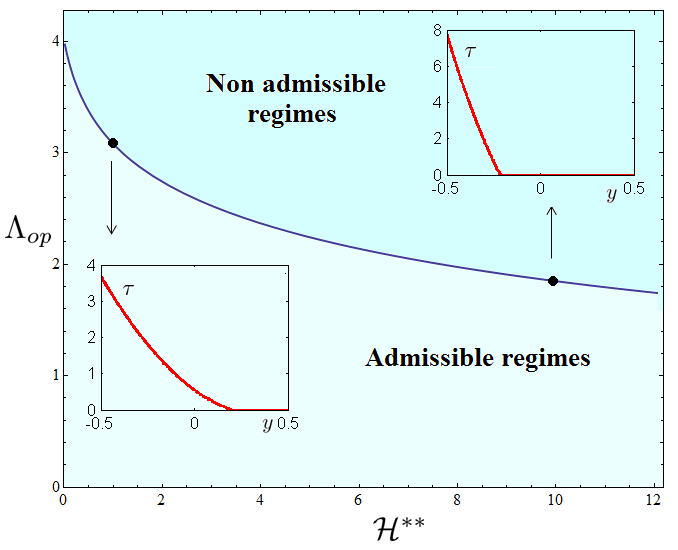}
\caption{\label{EffiOpt} Optimal efficiency as a function of the non mechanical cost $\mathcal{H}^{**}$. The lighter colored zone represents the admissible region and inserts show optimal configurations. Parameter $\mathcal{L}= 10.$ }
\end{center}
\end{figure}

The  optimal distribution $\tau(y)$ also depends  on the parameter
$\mathcal{H}^{**}=H ^{**}\sqrt{\eta \xi}/\bar{\tau}^2$ which contains
 condensed information about the mechanism of active force generation. Optimality
 here implies a trade-off between maximization of velocity $V$ and  minimization
 of the power of active stresses  $\mathcal{H}^{*}$. It is easy to show from
 (\ref{velstress}) that the maximum velocity  is $V^{\infty}=\mathcal{L}/2$ and
 that it corresponds to  full  localization  of motors at the trailing edge. Similarly, one can reach the lower bound  of the cost $\mathcal{H}^{*}=0$ by taking $\tau=1$,
 however, in this case $V=0$. The optimal trade-off depends on the value of
 $\mathcal{H}^{**}$ and  it is clear that optimally distributed motors  localize
 at $\mathcal{H}^{**}\rightarrow \infty $  and spread at $\mathcal{H}^{**}=0 $.

Mathematically, we have to solve an 'obstacle' problem for a quadratic
functional. Its general solution  has the form $\tau(y)=f(y)\theta(f(y))$, where
$f(y)=Ay^2+By+Cw$ and the constants $A,B,C$ can be found from a simple algebraic
minimization problem \cite{SI}. In the limit $\mathcal{H}^{**}\rightarrow 0$, we
obtain  $\Lambda \rightarrow (\mathcal{L}/2)\coth(\mathcal{L}/2)-1$ and
$\tau(y)\rightarrow 1-2y$. In the opposite limit $\mathcal{H}^{**}\rightarrow
\infty$  the efficiency tends to zero as $\Lambda \sim \mathcal{L}
(V^{\infty})^2/\mathcal{H}^{**}$ and $\tau(y)\rightarrow \delta(y+ 1/2)$ where $
\delta$ is the Dirac distribution.  In Fig.\ref{EffiOpt} we show the optimal
efficiency $\Lambda_{op}$ and some optimal profiles $\tau_{op}(y)$
for the intermediate values of $\mathcal{H}^{**}$. The regimes representing
physical "designs"  must be necessarily inside the admissible region bounded by
the optimal curve $\Lambda_{op}(\mathcal{H}^{**})$. Observe that under the assumption $\tau(y)\leq 0 $ we would have obtained exactly the same localization  of 'pushing' elements near the front end of the moving cell.

Suppose now that the active stress is generated by motors with mass density $\rho(x,t)$
and that $\tau=\chi \rho$, where $\chi$ is a positive material constant. Following
\cite{Bois2011, Recho2013a}, we assume that the transport of motors is governed by
the standard advection-diffusion equation which in dimensional variables takes the
form
\begin{equation}\label{stressc1}
\partial_t \rho+\partial_x(\rho v)-D\partial_{xx}\rho=0,
 \end{equation}
 where $D$ is the diffusion coefficient. After making the traveling wave ansatz,
 assuming no flux boundary conditions  and changing to dimensionless variables
we can integrate (\ref{stressc1}) to obtain the solution of this equation
 in the form \cite{Recho2013a}
\begin{equation}\label{stressc}
\rho(y)=\frac{e^{\lambda(\sigma(y)-V\mathcal{L}y)}}
{\int_{-1/2}^{1/2}e^{\lambda(\sigma(y)-V\mathcal{L}y)}dy}.
\end{equation}
Here density is normalized by $\bar{\rho}=\bar{\tau}/\chi$ and  $\lambda=\chi \bar{\rho}/(\xi D)$ is an additional parameter.

\begin{figure}[!h]
\begin{center}
\includegraphics[scale=0.7]{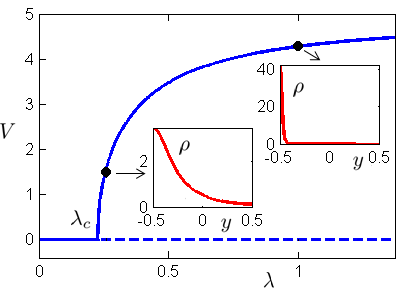}
\caption{\label{bifurstiff}  Cell velocity  as a function of  the nondimensional parameter $\lambda$. Inserts show the  density profiles for motors. The symmetric motile branch associated with negative velocities is not shown. Parameter $\mathcal{L}=10$. Characteristic scale of velocity is $v_c=\bar{\tau}/\sqrt{\eta\xi}\sim 0.02\mu m\cdot s^{-1}$}
\end{center}
\end{figure}

At small values of $\lambda$, the system of two equations (\ref{stressc}),
(\ref{velstress1}) has only a trivial solution   $\sigma(y)=\sigma_0=1$ and  $V=0$.
This solution becomes unstable at $\lambda_c=1-\omega^2/\mathcal{L}^2$ where
$\omega$ is the smallest root of the algebraic equation
$2\tanh(\omega/2)=\lambda_c\omega$. Through the pitchfork bifurcation shown in
Fig.\ref{bifurstiff} the cell becomes polarized and starts to move
\cite{Recho2013a}. As $\lambda$ increases the motors progressively concentrate at
the trailing edge. For  keratocytes we use \cite{Barnhart2011,Luo2012} to find
that $D \sim 10^{-13} m^2 \cdot s^{-1}$ which gives  $\lambda_e \sim 0.5$. We then
obtain an estimate $V_e=0.08\mu m s^{-1}$ which is very close to the value
measured in \cite{Verkhovsky1999}. Interestingly, for $\mathcal{L}\sim 10$ we get
$\lambda_c  \sim 0.23$ which implies that keratocytes operate rather close to the
bifurcation point. This near-criticality may carry considerable biological advantages since a cell
can easily switch from
static to motile state or change the direction of the already initiated motion.

The next step is to find the link between the  value of the non-dimensional parameter $\lambda$ fully characterizing the transportation problem and the cost parameter $\mathcal{H}^{**}$ from the optimization problem. For simplicity, we assume that the system is in contact with a thermal reservoir imposing a constant temperature $T$. To introduce the thermodynamic model (see \cite{SI} for more detail) we temporarily  bring back the dimensional variables.

Following the general theory of active gels \cite{Kruse2005} we describe the acto-
myosin network as a two-phase mixture with the total mass density $\hat{\rho}(x,t)$. It satisfies
the conservation equation $$\partial_t\hat{\rho}+\partial_x(\hat{\rho} v)=0,$$
which decouples from
the force balance problem due to the assumption of infinite compressibility:
if the velocity field $v(y)$ is known, $\hat{\rho}(y)$ can be reconstructed by standard methods
\cite{Recho2013,Recho2013a}.

The total free energy can be written as $$ F=\int_{-
L/2}^{L/2}\hat{\rho}fdx,$$ where $f(\phi,\zeta)$ is the energy density, which  depends
on the mass fraction (concentration) of motors in the mixture  $\phi(x,t)=\rho/\hat{\rho}$ and on a variable
$\zeta(x,t)$ characterizing the progress of a non-equilibrium  chemical reaction
(ATP hydrolysis) supplying energy to the motors. We can then write
\begin{equation}\label{stressc111}
H=-\dot{F},
\end{equation}
where the dot denotes the full time derivative. To compute the
right hand side we need to introduce the driving force of the reaction
 $A=-\partial_{\zeta} f>0$ which we assume to be fixed by an external "chemostat".  We also need to define the chemical potential of   motor
molecules   $\mu=\partial_{\phi} f$  whose nonzero gradient  is a crucial part of
the motility mechanism.   We can then write
 $$\dot{F}=\int_{-L/2}^{L/2} \hat{\rho} (A \dot{\zeta}+\mu \dot{\phi}) dx.$$ To show
 thermodynamic consistency of  (\ref{stressc1}) and to derive an additional
 equation for the variable $\zeta(x,t)$ we observe that the dissipation rate $R$
 can be written in the form $$R=W-\dot{F},$$ where $W=\int_{-
 L/2}^{L/2}\sigma\partial_xv dx$ is the external power. If  there
 are no sources of motors we can write $$\hat{\rho}\dot{\phi}=\partial_x J ,$$ where $J$
 is the diffusion flux.  Hence
 $$R=\int_{-L/2}^{L/2}\left( \sigma \partial_x
 v+\hat{\rho}\dot{\zeta}A+J\partial_x\mu\right) dx. $$
We postulate  linear relations between fluxes and forces:
$\sigma=l_{11}\partial_x v+l _{12}A $,
$\hat{\rho} \dot{\zeta}=-l_{21}\partial_x v+l_{22}A$ and
$J=l_{33}\partial_x\mu,$
where  for simplicity we omitted the
couplings between diffusion and reaction and between diffusion and viscosity.

Since motors are enzymes catalyzing the ATP
reaction we must deviate from the  Onsager scheme and assume that the
coefficients $l_{12}$ and $l_{22}$ are \emph{functions} of the motor density $\rho$.  To make the thermodynamic theory
consistent with our postulate $\tau=l _{12}A=\chi \rho$ we need to further
assume that these functions are linear. The other coefficients are assumed to be Onsagerian, for instance,
$l_{11}=\eta$. To compute the diffusion coefficient in (\ref{stressc1}) we notice
that for dilute mixtures  $\partial_{\phi}\mu=k_B T/\phi$ where
$k_B$ is Boltzmann's constant. If we make additional assumptions that the
variation of the total density is small $\partial_x\rho/\rho>>\partial_x\hat{\rho}/\hat{\rho}$ and the diffusion coefficient is concentration independent, we
recover  (\ref{stressc1}) with  $D=l_{33}k_B T/\bar{\rho}$ where $l_{33}$ is the mobility per unit volume. These assumptions clearly fail near the singularities of $\rho$ where the theory has to be appropriately modified.

An important outcome of our constitutive assumptions is an equation governing the reaction progress
\begin{equation}\label{stressc11}
\dot{\zeta}=\phi(bA-\frac{\chi}{A}\partial_xv),
 \end{equation}
where $b=\l_{22}/\rho $ is a constant parameter. Since the value of $A$ is  controlled from outside, (\ref{stressc11}) decouples from the rest of the system with $\zeta$ easily recoverable once the fields $v$ and $\phi$ are known.

We have now specified the force generation mechanism and can  use (\ref{stressc111}) to obtain an explicit
expression for the  cost function $H$. First, using the force balance equation we  write the mechanical cost function in the form
 $$H^{*}=\int_{-L/2}^{L/2} \left[ \xi v^2+\eta (\partial_xv)^2 \right] dx \geq 0,$$
where the two entries characterize contributions due to friction and viscosity.
The non-mechanical cost function can be written as
 $$H^{**}=\int_{-L/2}^{L/2} \left[ b\rho A^2+\frac{Dk_B T}{\bar{\rho}}(\partial_x \rho)^2
 \right] dx\geq 0.$$
Here the two terms are the cost of keeping the chemical reaction out of
equilibrium and the cost of supporting concentration gradients. Since the motion is driven exclusively by the "chemostat", we obtain for the physical solution that
both  $H^{*} \rightarrow 0$ and $H^{**} \rightarrow 0$ as $A \rightarrow 0$ \cite{SI}.

To make comparison with the optimization model we need to compute the dimensionless quantity
\begin{equation} \label{physhnm}
\mathcal{H}^{**}=\mathcal{M}\mathcal{L}+\frac{\mathcal{E}}{\lambda \mathcal{L}}\int_{-1/2}^{1/2}(\partial_y\rho)^2dy .
\end{equation}
where we introduced two new  parameters of the problem: $\mathcal{M}=\eta
bA^2/(\bar{\rho}\chi^2)$ and $\mathcal{E}=k_BT/\chi$. If motors operate in stall
conditions and form  bipolar contractile units with size $d$, the produced  force
is $p \sim 2\chi/d$.  For myosin II we have  $d \sim 0.15\mu m$ and  $p \sim 1.5
pN$ \cite{Howard2001} which gives $\chi\sim 1.1\times 10^{-19} J$ and  using $k_BT
\sim 4.3\times 10^{-21} J$ we   obtain  $\mathcal{E}\sim 0.04$. Notice also that
$bA^2$ is the free energy consumption rate per motor and therefore that
it is equal to
$kA$ where $k$ is the rate of ATP turnover per motor. Since $A\sim 25 k_BT$ and
$k\sim 25s^{-1}$ \cite{Howard2001}, we obtain $\mathcal{M}\sim 0.053.$
\begin{figure}[!h]
\begin{center}
\includegraphics[scale=0.55]{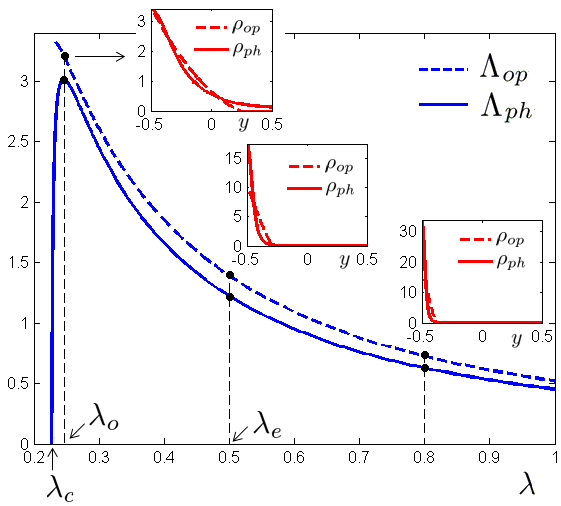}
\caption{\label{Realeff2} Physical and optimal efficiencies as functions of the dimensionless parameter $\lambda$. Inserts compare optimal and physical distributions of motors. Parameters: $\mathcal{L}=10$, $\mathcal{M}= 0.053$ and $\mathcal{E}= 0.04 $.}
\end{center}
\end{figure}

We can now the fix the parameters $\mathcal{E},\mathcal{M}$ and compare physical and optimal
efficiencies at different values of the remaining parameter $\lambda$. The
efficiency in the physical model is $ \Lambda_{ph}(\rho_{ph}, \mathcal{H}^{**})$
where $\rho_{ph}(y,\lambda)$ is the solution of (\ref{stressc1}) and
$\mathcal{H}^{**}(\lambda)$ is taken from (\ref{physhnm}). The ensuing function
$\Lambda_{ph} (\lambda)$  is to be compared with the optimal efficiency
$\Lambda_{op} (\lambda)=\Lambda_{op}(\rho_{op},\mathcal{H}^{**})$ where $\rho_{op}(y,
\mathcal{H}^{**})$ is the solution of the optimization problem and
$\mathcal{H}^{**}(\lambda)$ is the same as above.
\begin{figure}[!h]
\begin{center}
\includegraphics[scale=0.6]{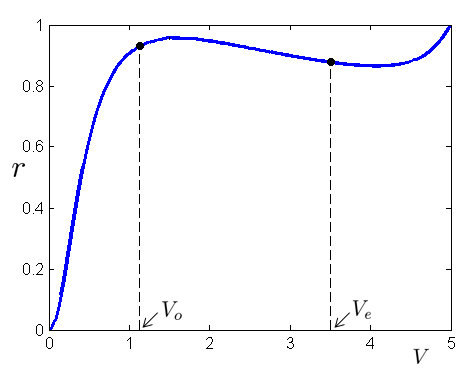}
\caption{\label{sensitivity} Ratio of physical and optimal efficiencies as a function of cell velocity. Indicated are the low bound of the regimes of near-optimal performance $V_o (\lambda_o)$ and the experimentally observed regime $V_e (\lambda_e)$. The symmetric branch associated with negative velocities is not shown. Parameters: $\mathcal{L}=10$, $\mathcal{M}= 0.053$ and $\mathcal{E}= 0.04 $. Characteristic scale of velocity is $v_c=\bar{\tau}/\sqrt{\eta\xi}\sim 0.02\mu m\cdot s^{-1}$}
\end{center}
\end{figure}

The results of the comparison are summarized in Fig.\ref{Realeff2}. The function
$\Lambda_{ph}(\lambda)$  displays a single maximum at $\lambda_o \sim 0.24 $. For
all $\lambda \geq \lambda_o$, the physical model remains close to the optimal one
which is crucial since at $\lambda \rightarrow \infty$ the rate of energy
consumption diverges  $\mathcal{H}^{*}\sim \mathcal{L}^3\lambda/3,
\mathcal{H}^{**}\sim \mathcal{E}\mathcal{L}^5\lambda^2/15$. In this (high
velocity) limit both physical and optimization problems  generate the same profiles
with  motors infinitely localized at the trailing edge of the moving cell. The
robust  optimality in the range  $V \geq V_0 (\lambda_o) \sim 1.1$  is illustrated
further in Fig. \ref{sensitivity} where we show the ratio
$r(V)=\Lambda_{ph}/\Lambda_{op}\leq 1$. The presence of a quasi-plateau on this graph in
the biologically relevant range of velocities \cite{Verkhovsky1999} and the fact
that in this range the physical and the optimal efficiencies are close
suggest that
the system is tuned to optimality. In the immediate vicinity of the motility initiation point $V=0$
where  $\lambda \sim \lambda_c$, the asymptotic solutions in the two models have the same general structure $\rho(y)-1 \sim  (\lambda-\lambda_c)^\kappa f(y)$, however,  since   $\kappa_{ph}=1/2 $  and  $\kappa_{op}=0$ the propulsion machinery operates sub-optimally in this regime and this may explain
why small velocities have not been observed in experiments.

In conclusion, we have shown that in \emph{contraction-dominated} crawling the optimal
trade-off between Stokes performance  and the metabolic cost is achieved by
localization  of contractile units at the trailing edge of the moving cell. A
simple advection-diffusion model of motor redistribution based on the active gel
theory performs almost optimally in the range of parameters suggested by in vivo
measurements. The fact that the near-optimal behavior is \emph{robust} and extends
into the domain of parameters where  sub-optimality would be particularly costly,
suggests that contraction-dominated crawling presents an example of a remarkably perfected biological
mechanism.

The authors thank  F. Alouges, G. Geymonat and T. Putelat for helpful discussions. P.R. thanks Fondation Pierre-Gilles de Gennes for generous support.

\end{document}


\maketitle

\newtheorem{theorem}{Result}
\newtheorem{definition}{Definition}
\newtheorem{proposition}{Proposition}
\newtheorem{remarque}{Remark}
\newtheorem{preuve}{Formal proof}
\newtheorem{exemple}{\emph{Example}}
\newcommand{\bs}{\left\{}
 \newcommand{\es}{\right.}
 \newcommand{\ba}{\begin{array}}
 \newcommand{\ea}{\end{array}}
 \newcommand{\be}{\begin{equation}}
 \newcommand{\ee}{\end{equation}}
 \newcommand{\Div}{\nabla.}
 \newcommand{\ds}{\displaystyle}

\newcommand{\fonction}[5]{\begin{array}{l|rcl}
#1: & #2 & \longrightarrow & #3 \\
    & #4 & \longmapsto & #5 \end{array}}

\renewcommand{\theequation}{\Roman{equation}}
\renewcommand{\thefigure}{\Roman{figure}}

\section{Solution of the optimization problem}
Here we show how the infinite dimensional optimization problem formulated in the
main text can be reduced to an algebraic  minimization problem in two dimensions.

If we substitute the function $\sigma(y)$ from (2) into the dimensionless expression of efficiency, we obtain:
\begin{equation}\label{effifirst}
\Lambda=\frac{\mathcal{L}V^2}{\mathcal{L}\int_{-1/2}^{1/2} \tau(y) ^2dy-2\sigma_0^2\tanh(\frac{\mathcal{L}}{2})-\mathcal{L}^2\int_{-1/2}^{1/2}\int_{-1/2}^{1/2}\Psi(y,v)\tau(y)\tau(v)dydv +\mathcal{H}^{**}}.
\end{equation}
The constants $V$ and $\sigma_0$ are defined in terms of the unknown function
$\tau(y)$  in Eq.(3) of the main text (see also   Eq.(III) below).  Our strategy is to first
fix $V$ and $\sigma_0$, optimize (\ref{effifirst}) with respect to $\tau(y)$
and then to optimize the result
with respect to $V$ and
$\sigma_0$.

The  first problem is equivalent to minimizing the denominator in (\ref{effifirst})
\begin{equation}\label{minipb}
\underset{\tau}{\text{min }} \left[Q(\tau)=\int_{-1/2}^{1/2}\tau(y)^2dy-\mathcal{L}\int_{-1/2}^{1/2}\int_{-1/2}^{1/2}\Psi(y,v)\tau(y)\tau(v)dydv\right],
\end{equation}
under the constraints
\begin{equation}\label{constraints}
\left\{ \begin{array}{c}
2V\sinh(\frac{\mathcal{L}}{2})=-\mathcal{L}\int_{-1/2}^{1/2}\sinh(\mathcal{L}y)\tau(y)dy \\
2\sigma_0\sinh(\frac{\mathcal{L}}{2})=\mathcal{L}\int_{-1/2}^{1/2}\cosh(\mathcal{L}y)\tau(y)dy\\
\int_{-1/2}^{1/2}\tau(y)dy=1.
\end{array} \right.
\end{equation}
An additional constraint $\tau(y)\geq 0$ states
the active forces in our system are contractile.

In order to take the constraints into account, we introduce three scalar Lagrange
multipliers $(\kappa_0,\kappa_1,\kappa_2)$ and  a non negative function
$\kappa(y)$. Then the condition that the efficiency is maximal takes
the form
\begin{equation}\label{fredholm}
(I-Q)\tau(y)=\Gamma(y)
\end{equation}
where
$$Q \tau=\mathcal{L}\int_{-1/2}^{1/2}\Psi(y,v)\tau(v)dv$$
and
$$\Gamma(y)=\kappa_0-\kappa_1 \sinh(\mathcal{L}y)+\kappa_2
\cosh(\mathcal{L}y)+\kappa(y).$$
We invert the kernel of equation (\ref{fredholm}) using an expansion in
eigenfunctions \cite{Mikhlin1960} to obtain
\begin{equation}\label{repregene}
\tau(y)=(I-Q)^{-1}\Gamma(y)=\int_{-1/2}^{1/2}\phi(y,v)\Gamma(v)dv.
\end{equation}
Here
\begin{equation}\label{solkern}
\phi=\delta(y-v)+\mathcal{L}^2\left[(\frac{1}{2}+y)(\frac{1}{2}-v)\theta(v-y)+(\frac{1}{2}+v)(\frac{1}{2}-y)\theta(y-v)\right],
\end{equation}
$\theta$ is the Heaviside function and $\delta$ is the Dirac distribution. Using the  expression for $\Gamma$ we obtain
\begin{equation}\label{solsigncons}
\tau(y)=P(y)+\int_{-1/2}^{1/2}\phi(y,v)\kappa(v)dv.
\end{equation}
The function
\begin{equation}\label{poly}
P(y)=\kappa_0\psi_0(y)+\kappa_1\psi_1(y)+\kappa_2\psi_2(y)
\end{equation}
is a parabola since
\begin{equation}\label{basisP}
\psi_0(y)=1-\frac{\mathcal{L}^2}{2}( y+\frac{1}{2})( y-\frac{1}{2})\text{, }\psi_1(y)=-2y\sinh(\frac{\mathcal{L}}{2})\text{ and  } \psi_2(y)=\cosh(\frac{\mathcal{L}}{2}).
\end{equation}
The Lagrange multipliers can be found from Karush-Kuhn-Tucker conditions \cite{Boyd2004}
\begin{equation}\label{Lagmultisyst}
 \begin{array}{c}
1=\kappa_0 A_0+\kappa_2 A_2+\int_{-1/2}^{1/2}\kappa(y)\psi_0(y)dy\\
2V\frac{\sinh(\frac{\mathcal{L}}{2})}{\mathcal{L}}=\kappa_1 S_1+\int_{-1/2}^{1/2}\kappa(y)\psi_1(y)dy\\
2\sigma_0\frac{\sinh(\frac{\mathcal{L}}{2})}{\mathcal{L}}=\kappa_0 C_0+\kappa_2 C_2+\int_{-1/2}^{1/2}\kappa(y)\psi_2(y)dy\\
\tau(y)\geq 0 \text{, } \kappa(y)\geq 0 \text{ and } \kappa(y)\tau(y)=0,
\end{array}.
\end{equation}
where
$$
 \begin{array}{c}
A_0= 1+\frac{\mathcal{L}^2}{12}\text{, } A_2= \cosh(\frac{\mathcal{L}}{2})\text{, } \\
S_1= \frac{2(1-\cosh(\mathcal{L}))+\mathcal{L}\sinh(\mathcal{L})}{\mathcal{L}^2}\text{, }\\
C_0= \cosh(\frac{\mathcal{L}}{2})\text{ and }C_2= \frac{\sinh(\mathcal{L})}{\mathcal{L}}.
\end{array}
$$
If the function $\kappa(y)$ is known, the Lagrange multipliers $(\kappa_0,\kappa_1,\kappa_2)$ are readily found from the system of linear equations (\ref{Lagmultisyst}). To find $\kappa(y)$  we first notice that the function
\begin{equation}\label{formg}
\tau(y)=P^+(y),
\end{equation}
where $P^+(y)=\text{\text{max }}(0,P(y))$  satisfies the constraint $\tau(y)\geq 0$. The  associated $\kappa(y)$  can be obtained by inverting (\ref{solsigncons})
\begin{equation}\label{kappasol}
\kappa(y)=(I-Q)P^-(y).
\end{equation}
where $P^-(y)=\text{\text{max }}(0,-P(y))$.

Next we show that such  $\kappa(y)$  satisfies the last set of conditions in (\ref{Lagmultisyst}).
Define the function
 $\sigma_{-}(y)=QP^-(y)$
 which solves the boundary value problem,
\begin{equation}\label{BVPPm}
-\mathcal{L}^{-2}\sigma_{-}''+\sigma_{-}=P^-\geq 0\text{, }\sigma_{-}(\pm1/2)=0.
\end{equation}
Maximum principle \cite{Gilbarg2001} ensures that $\sigma_{-}(y)\leq P^-(y) $ and thus  $\kappa(y) \geq 0$. We can also see that
 $\tau(y)\kappa(y)=-P^+(y)\sigma_{-}(y)=0,$
since whenever $P^+(y)>0$, the function $\sigma_{-}$ satisfies  (\ref{BVPPm}) with zero Dirichlet boundary conditions. Thus (\ref{formg}) satisfies all the required conditions.

The resulting optimal distribution of active stresses  is
  \begin{equation}\label{BVPPm1}
\tau(y)=(Ay^2+By+C)\theta(Ay^2+By+C).
 \end{equation}
Notice that this function has a singularity at a point where $Ay^2+By+C=0$.
Instead of  expressing  the constants $A,B,C$ in terms of $V$ and $\sigma_0$ and
then optimizing the efficiency with respect to these two variables, in our numerical code we directly
minimize efficiency with respect to $A,B$.  The third constant  $C$ is determined by  the
constraint $\int_{-1/2}^{1/2}\tau(y)dy=1.$

As an example, consider the limiting problem with  $\mathcal{H}^{**}=0$. Suppose
first that there is no sign constraint on $\tau(y)$ and denote the corresponding
optimal distribution $\tilde{\tau}(y)$.  Then $\kappa\equiv 0$ and the system
(\ref{Lagmultisyst}) is linear which allows one to find $(\kappa_0, \kappa_1,
\kappa_2)$ explicitly as  functions of $\sigma_0$ and $V$. We  obtain
\begin{equation}\label{L1}
\tilde{\Lambda}(V,\sigma_0)=\frac{V^2}{\frac{V^2}{\mu(\mathcal{L})}
+\frac{(1-\sigma_0)^2}{\frac{\mathcal{L}^2}{12}+1-\frac{\mathcal{L}}{2}
\tanh(\frac{\mathcal{L}}{2})}}.
\end{equation}
As all terms in (\ref{L1})  are positive it is clear that,
$$\tilde{\Lambda}(V,\sigma_0)\leq \mu(\mathcal{L})=\frac{\mathcal{L}}{2}\coth(\frac{\mathcal{L}}{2})-1.$$
We now get back to the initial problem with the sign constraint. Since  by
definition $\Lambda\leq\tilde{\Lambda},$  we can write $\Lambda\leq
\mu(\mathcal{L})$. It is easy to find a non negative function $\tau$ from the
family (\ref{BVPPm1}) which saturates the bound. A simple substitution shows that
for $\tau(y)=1+\alpha y $ with  $\alpha \in[-2,2]$ one obtains
$\Lambda = \mu(\mathcal{L})$. This means that the whole one parametric family is
optimal. Negative (positive) values of $\alpha$ correspond to positive (negative)
velocities. Therefore, the optimal velocities range is between
$\pm2\mu(\mathcal{L})/\mathcal{L}$.
From this set
only configurations with $\alpha=\pm 2$ can be recovered  in the limit $\mathcal{H}^{**}\rightarrow 0$  from the sequence of optimal configurations with $\mathcal{H}^{**}>0$.

\section{Energetic cost of maintaining a steady state}

We first specialize some standard relations of continuum thermodynamics of nonequilibrium process for our problem \cite{DGM, Kruse2005}.  Then we introduce the crucial definition for the 'rate of free energy consumption' and  compute its value for our traveling wave solution.

Recall, that our finite 1D layer of reacting viscous fluid is exposed to:
(i)  distributed (bulk) forces  $-\xi v$ due too friction  and (ii)   surface
tractions $\sigma_0$ on the boundaries $x=-L/2,L/2$ due to the cortex. The power of these
\emph{external} forces can be written as
\begin{multline}\nonumber
\dot{W}= -\int_{-L/2}^{L/2}\xi v^2 dx+\sigma_0 (v(L/2)-v(-L/2)) \\
= \int_{-L/2}^{L/2}(-\xi v^2+\partial_x(\sigma v))dx
= \int_{-L/2}^{L/2}(-\xi v^2+v\partial_x\sigma+\sigma\partial_x v)dx.
\end{multline}
By taking into account the force balance
$$\partial_x\sigma=\xi v,$$
we  can further rewrite $\dot{W}$ as the power of the \emph{internal} forces
$$ \dot{W}= \int_{-L/2}^{L/2}\sigma\partial_xv dx.$$

The next step is to compute the rate of change of the free energy
$$F=\int_{-L/2}^{L/2} \hat{\rho} fdx,$$
where $\hat{\rho}$ is the total density of the mixture which is a conserved quantity
$$\partial_t\hat{\rho}+\partial_x(\hat{\rho} v)=0.$$ In addition to temperature, the free energy density may depend on $\hat{\rho}$, on the mass fraction of the motor component of the mixture $\phi=\rho/\hat{\rho}$ and on the  advancement of the hydrolysis
reaction per unit of mass $\zeta$.  Due to the assumption of infinite compressibility and the presence of a thermostat, we are left with only  two essential variables, so
 $$f=f(\phi,\zeta).$$ 
Hence we can write
$$\dot{F}=\int_{-L/2}^{L/2} \hat{\rho}(A \dot{\zeta}+\mu \dot{\phi})dx$$
where
$$A(\phi,\zeta)=-\partial_\xi f$$ is the affinity of the reaction and $$\mu(\phi,\zeta)=\partial_{\phi} f$$ is the chemical potential of the motors.

Finally, we make  an assumption that  motors are not created in the bulk by writing
\begin{equation}\label{diff}
\rho_0\dot{\phi}=\partial_x J,
\end{equation}
where $J$ is the flux of  motors.

For our isothermal system the rate of irreversible entropy production can be written as
$$T\dot{S}_{i}= \dot{W}-\dot{F}\geq 0. $$
Since there is no fluxes on the boundaries,  we obtain
$$T\dot{S}_{i}= \int_{-L/2}^{L/2}(\sigma \partial_x v+\hat{\rho}\dot{\zeta}A+J\partial_x\mu)dx\geq 0. $$
The three terms in the right hand side can be interpreted as  products of the
thermodynamic fluxes $\sigma, \hat{\rho}\dot{\zeta}, J$ and  the conjugate
thermodynamic forces $\partial_x v, A,  \partial_x\mu$. We make a simplifying
assumption that fluxes and forces are related through
Onsager type relations
\begin{equation}\label{forceflux}
\begin{array}{c}
\sigma=l_{11}\partial_x v+l_{12}A +l_{13}\partial_x\mu\\
\hat{\rho}\dot{\zeta}=l_{21}\partial_x v+l_{22}A +l_{23}\partial_x\mu\\
J=l_{31}\partial_x v+l_{32}A +l_{33}\partial_x\mu\\
\end{array}
\end{equation}
Here the different tensorial nature of the fluxes/forces is not an issue because
the anisotropy is prescribed by our 1D ansatz.   

Finally, we   make another simplifying
assumption  that the diffusion flux $J$  depends only on $\partial_x\mu$ which
implies that $l_{31}=l_{32}=l_{13}=l_{23}=0$. Since time inversion symmetry
requires that  $l_{12}=-l_{21}$  we are left with four coefficients
$l_{11},l_{22},l_{33}, l_{12}$.

We assume that two of these coefficients describe genuinely linear dissipative mechanisms and are therefore standard: $l_{11} =\eta \geq 0$ is the
viscosity  and $l_{33} \geq 0$ is a mobility per unit volume. To specify the diffusion coefficient fully we first rewrite (\ref{diff}) in the form
$$\partial_t\rho+\partial_x(\rho v)=\partial_x(l_{33}\partial_x\mu).$$
Assuming that the acto-myosin gel is  a dilute mixture we can write
$$f=f_0(\zeta)+k_BT \phi \log\phi$$
where $k_B$ is the Boltzmann constant. Therefore
$$\mu=\mu_0+k_BT\log\phi$$
and
$$\partial_x\mu=k_BT\left( \frac{\partial_x\rho}{\rho}-\frac{\partial_x\hat{\rho}}{\hat{\rho}}\right).$$
To recover a standard diffusion equation we need to make an additional assumption that  the variation of the total density is small compared to the variation of the density of  motors
$$\frac{\partial_x\rho}{\rho}>>\frac{\partial_x\hat{\rho}}{\hat{\rho}}.$$
Then we obtain the Einstein-Smoluchowski relation
$$D=\nu k_BT,$$
where $\nu=l_{33}/\rho$ is the mobility per motor.  To remain in the framework of  Onsager theory  we need to assume that $\rho\sim \bar{\rho}$ and  $l_{33}=l_{33}(\bar{\rho})$;   this  approximation clearly fails near the singularities of $\rho$ where the model needs to be appropriately modified. 

To finalize the model we need to specify the two remaining coefficients:  $l_{12}$, which describes chemo-mechanical coupling \cite{Finlayson1969} and does not contribute to entropy production and $l_{22}$, which
describes reaction kinetics and must be non-negative to ensure positive definiteness of the dissipation. Notice that in this model we   deal  with an enzymatic reaction. This is
a nonlinear phenomenon because the kinetics is accelerated in the presence of motors. Therefore the straightforward linear
Onsager relations do not apply and we need to replace them with quasi-linear relations  by  making Onsager
coefficients dependent on the fields. The simplest way to take 
the enzymatic activity of the motors into account
is to assume that
$$l_{12}=a \rho, l_{22}=b \rho,$$
where $a,b$ are now constants. One consequence of these assumptions is the constitutive relation for stress
$$\sigma=\eta\partial_x v+aA\rho,$$
where the first term describes classical viscosity while the second
term represents the active stress due to mechano-chemical coupling.  We assume that $a\geq0$ which ensures that the reaction induced stresses are contractile whenever $A>0$. Another consequence of our quasi-linearity assumption is the specific form of the kinetic equation for the hydrolysis
reaction
$$\partial_t(\hat{\rho} \zeta)+\partial_x(\hat{\rho} \zeta v)=\rho(bA-a\partial_xv).$$
Observe that in this model the reaction stops completely in the absence of motors ($\rho=0$ ).  

If the passive system described above is left isolated, it reaches equilibrium
(all fluxes vanish and the entropy production stops). To maintain the non-
equilibrium state, the dissipated energy  must be continuously
 re-injected into the system. Since the temperature reservoir  is
 in equilibrium, the system is not exchanging mass with the environment, and no directional forces conduct external work, the
only way to prevent the equilibration of the system, is to keep the driving
force of the reaction $A$  away from zero.

More specifically, this means that the corresponding ratio of the concentrations of ATP, ADP and P is kept at a fixed 'distance'  from its equilibrium value through incessant breaking or assembling of the associated molecular complexes.
The exact microscopic mechanism of such a continuous 'fine tuning' performed by an
\emph{external} 'chemostat' is not fully clear,  however, in  our formalism such assumption of perpetual disequilibrium is tantamount to the assumption that
$$f_0(\zeta)=-A\zeta,$$
 where $A>0$ is a prescribed constant.
This bottomless decrease of the energy landscape mimics the continuous rebuilding of the non-equilibrium state despite the tendency of the system to reach equilibrium (where $A=0$). The crucial assumption that $A=const$ allows one to decouple the reaction equation from the system and compute the  energetic cost of maintaining disequilibrium  from  the knowledge of the free energy losses per unit time that must be compensated externally.

We now make the important assumption that the cost of maintaining the non-equilibrium steady state is equal to the rate of consumptions by the system of its free energy 'reserves' (that are being continuously replenished)
$$H=H^*+H^{**}=-\dot{F}.$$
Then, using the constitutive relations we can write
\begin{multline}\label{foursinks1} 
-\dot{F}=\int_{-L/2}^{L/2} (\rho_0\dot{\zeta}A+J\partial_x\mu)dx \\
=\int_{-L/2}^{L/2} (-\chi \rho\partial_xv+bA^2\rho+D\frac{k_BT}{\bar{\rho}}(\partial_x \rho)^2)dx.
\end{multline}
If we multiply the force balance equation by $v$ and use the boundary conditions, we obtain
$$
-\chi\int_{-L/2}^{L/2}\rho \partial_xvdx= \xi\int_{-L/2}^{L/2} v^2dx+ \eta\int_{-L/2}^{L/2}(\partial_xv)^2dx.
$$
By substituting this  relation into (\ref{foursinks1}) we finally obtain
\begin{equation}\label{foursinks}
-\dot{F}=\int_{-L/2}^{L/2}(\xi v^2+\eta (\partial_xv)^2+bA^2\rho+D\frac{k_BT}{\bar{\rho}}(\partial_x \rho)^2)dx\geq 0.
\end{equation}
We can now identify the terms $H^*$ and  $H^{**}$. The mechanical cost function
$$H^*=\xi \int_{-L/2}^{L/2} v^2 dx+\eta \int_{-L/2}^{L/2}  (\partial_xv)^2 dx \geq 0$$
is a sum of contributions due to frictional and viscous
dissipation. The non-mechanical part
$$H^{**}=bA^2\int_{-L/2}^{L/2} \rho dx+D\frac{k_BT}{\bar{\rho}}\int_{-L/2}^{L/2}(\partial_x \rho)^2dx.$$
represents  the   energetic cost  of maintaining
the finite rate of chemical reaction and the cost of keeping a nonzero concentration
gradient.

To summarize, the steady state self-propulsion in the proposed model requires: (i) work against friction
which is necessary for acquiring momentum, (ii)  work against viscosity which is a
mechanism of long range interactions in the cell providing mechanical coordination at distant points,
(iii) work against diffusion to ensure optimal distribution of motors and finally, (iv) work
to keep the reaction 'burning' which ensures mechano-chemical generation of active
forces. All these processes are dissipative and the consumed free energy  needs to be compensated.

As we have seen, the steady state is maintained  due to the exterior chemostat which ensures
that $A\neq 0$. If $A=0$, the active stress is equal to zero and the velocity
vanishes so the first two terms in (\ref{foursinks}) contributing to the cost vanish as well. The third
term obviously vanishes because it is proportional to $A^2$. In the absence of
flow, the density of motors becomes  homogeneous because the destabilizing
advection disappears. Therefore  the fourth term is also equal to zero.
\begin{figure}[!h]
\begin{center}
\includegraphics[scale=0.6]{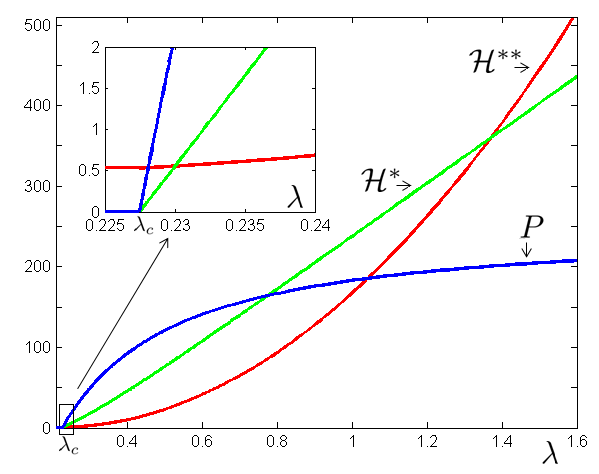}
\caption{\label{entrieseff}  Dimensionless Stokes power $P=\mathcal{L}V^2$ and metabolic costs $\mathcal{H}^*$ and $\mathcal{H}^{**}$ as functions of $\lambda$. Parameters are $\mathcal{L}=10$, $\mathcal{M}=0.053$ and $\mathcal{E}=0.05$.}
\end{center}
\end{figure}

To illustrate this argument, we present in Fig. \ref{entrieseff}  different terms entering the expression for efficiency
$$\Lambda=\frac{P}{\mathcal{H}^*+\mathcal{H}^{**}}$$
as functions of the parameter  $\lambda$  which can be viewed as a dimensionless version of $A$. One can see that at $A=0$ all three terms are equal to zero. It is also easy to show by asymptotic expansion that right above the motility initiation threshold $A = A_c$ the mechanical energy rates ($\mathcal{H}^{*}$ and  $P$) depend linearly on $A-A_c$ while the non-mechanical cost function $\mathcal{H}^{**} \sim A_c^2$ , see the inset in Fig. \ref{entrieseff}.